# Optical ultracompact directional antenna based on a dimer nanorod structure

*Fangjia Zhu[1*], María Sanz-Paz[1], Antonio Fernández-Domínguez[2], Mauricio Pilo-Pais[1] and Guillermo P. Acuna[1*]*

[1] Department of Physics, University of Fribourg, Chemin du Musée 3, Fribourg CH-1700, Switzerland.

[2] Departamento de Física Teórica de la Materia Condensada and Condensed Matter Physics Center (IFIMAC), Universidad Autónoma de Madrid, E-28049 Madrid, Spain.

**Corresponding Authors**

* E-mail: fangjia.zhu@unifr.ch (F.Z.), guillermo.acuna@unifr.ch (G.P.A.)

**Abstract**

Controlling directionality of optical emitters is of utmost importance for their application in communication and biosensing devices. Metallic nanoantennas have been proven to affect both excitation and emission properties of nearby emitters, including directionality of their emission. In this regard, optical directional nanoantennas based on a Yagi-Uda design have been demonstrated in the visible range. Despite this impressive proof of concept, their overall size ($\sim\lambda^2/4$) and considerable number of elements represent obstacles for the exploitation of these antennas in nanophotonic applications and for their incorporation onto photonic chips. In order to address these challenges, we investigate an alternative design. In particular, we numerically demonstrate unidirectionality of an "ultracompact" optical antenna based on two parallel gold nanorods (side-by-side dimer). Our results show that exciting the antiphase mode by an emitter placed in the near-field can lead to unidirectional emission. Furthermore, in order to verify the feasibility of this design, we study the effect on the directionality of several parameters such as shape of the nanorods, possible defects in dimer assembly, and different position and orientation of the emitter. We conclude that this design is robust to changes, making it experimentally achievable.

**Introduction**

The combination of optical single emitters with nanoantennas holds great promise for quantum information applications[1] and for their integration on photonic chips[2]. However, to achieve this, not only ultrabright[3] but also directional single photon sources[4, 5] are required. Optical nanoantennas act as effective bridges between receivers and transmitters, and as such they have been widely applied for manipulating interactions between light and matter.[6] To date, several schemes have been used to affect emitter properties, including tuning excitation[7], decay rate[8], polarization[9], frequency conversion[10, 11], spectral modulation[12], nonlinear processes[13] and emission direction[14, 15]. The most commonly used design for directional emission is based on the Yagi-Uda structure[14-17] inspired from radio frequency devices. There are also other designs proposed to achieve directional emission or scattering in the visible range, ranging from a pair of bimetallic nanodisks[18, 19] to V-antennas[20, 21], trimers[22] or a nanorod standing on a disk[23]. However, large metal surfaces of these antennas may introduce high absorption losses and the accompanying Joule heating will cause dysfunction of nearby temperature dependent devices on the photonic chips[24]. For example, the Yagi-Uda antenna is based on the far-field interference between the electromagnetic waves produced by a feed, a reflector and some directors, leading to a footprint of about $\lambda^2/4$ due to the number of elements and the specific gaps between them that are required[25]. Furthermore, these geometric constraints and the precise emitter positioning that is needed requires demanding and serial top-down fabrication techniques, such as electron or ion beam lithography[16, 26], which are not accessible to common chemical laboratories.

All these shortcomings call for an alternative compact antenna design for the integration of directional emitters into photonic chips. Pakizeh *et al.* have theoretically proposed an ultracompact directional antenna design, which is based on a stacked nanodisk dimer[27]. In this case, unidirectionality is achieved by exciting the antiphase plasmon mode through a localized emitter[28]. Similarly, Shen *et al.* used nanostrip dimers embedded on a dielectric material to simulate unidirectionality[29] and achieved a compact plasmonic-diamond hybrid nanostructure[30]. On the other hand, Bonod *et al*. have proposed a different ultracompact directional antenna design, which is based on two coupled nanospheres[31]. Both structures achieve unidirectionality by adjusting phase differences, introduced by mode hybridization or optical path difference, respectively. At the same time, other theoretical designs based on dielectric or hybrid nanostructures[32], phase-change materials[33] or plasmonic structures supporting magnetic modes[34] achieve unidirectional emission by using electric and magnetic dipoles interference to meet the Kerker condition[35]. Still, to date experimental studies on compact directional optical antennas addressing single emitters are limited.

Here, we propose a directional ultracompact antenna design based on two parallel gold nanorods (AuNRs), following the basis of Pakizeh's scheme[27]. We present a numerical study of the behaviour of these antennas using Finite Element Method (FEM) simulations based on AuNRs of specific size and shape. In order to study the feasibility of this design, we took several factors into consideration, e.g., common fabrication limitations like nanoparticle´s commercial dimensions, simplicity (reducing the number of needed elements), coupling to emitters and optimization of footprint. We also considered the effect of various geometrical parameters such as the gap between AuNRs, or the position and orientation of the emitter. By converting far-field signal to back focal plane (BFP) images, we quantified unidirectionality of antennas with forward to backward ratio ($F/B$). Finally, we used an analytical two dipole model to explain the mechanism behind the observed results, and quantified the phase difference between the two AuNRs. Overall, the proposed ultracompact antenna design made of a nanorod dimer and an adjacent emitter shows excellent and robust unidirectionality, with a $F/B$ value that can be as large as 14.2 dB.

**Results**

The main parameters considered for simulating the behaviour of the ultracompact antenna are depicted in Fig. 1. Two parallel AuNRs form a dimer in a side-by-side configuration. A dipole is positioned above the tip of one of the AuNRs at a distance gap1, and the two nanorods are separated by a distance gap2 between them. These three elements (two AuNRs and a dipole emitter) make up the ultracompact nanorod dimer antenna (NRDA) studied in this work. For comparison, we also study a system with a single nanorod with a dipole emitter coupled to its tip (i.e. without the right nanorod in Fig. 1(a)). This structure is hereafter referred to as nanorod monomer antenna (NRMA). We also used glass as a substrate, in agreement with typical experimental conditions. The distance from the AuNRs to the glass surface is gap3, as shown in Fig. 1(b).

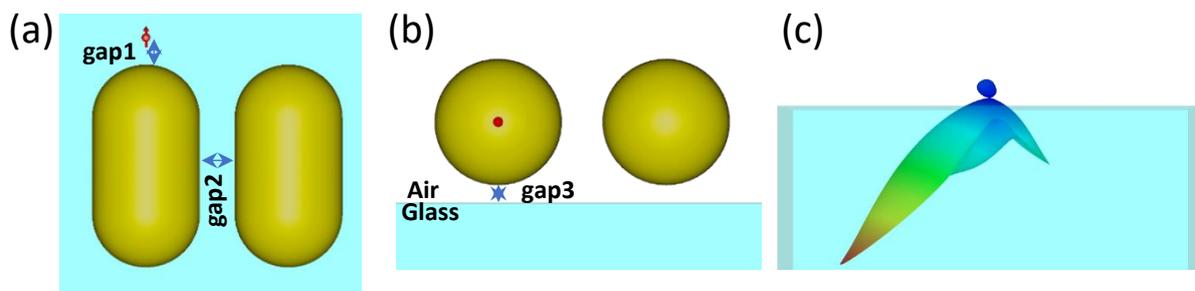

*Figure 1 (a, b) Sketches of the ultracompact nanoantenna based on two nanorods and a single dipole emitter (red arrow) on a glass substrate. (c) Corresponding radiation pattern when gap1, gap2 and gap3 are set to 5 nm.*

For the initial FEM far-field simulations, we chose AuNRs with commercial sizes: 40 nm diameter, 68 nm length and ideal semi-sphere cap. Distances gap1, gap2 and gap3 were set to 5 nm, and the background medium is set to vacuum (n=1). Unless otherwise specified, we used these parameters for all FEM simulations. Such simulations show that the radiation pattern of NRDA happens to be asymmetric within a specific wavelength range (Fig. 1(c)), with the main emission lobe occurring at the side of the antenna where the emitter is placed. As will be later discussed, the wavelength range where directionality occurs corresponds to the antiphase plasmon mode of the NRDA.

For better visualization, and to better match the results with what it is commonly measured experimentally, we translate this 3D far-field emission pattern into 2D BFP images. This is done by introducing an optical lens that projects the Fourier transform by converting every $\theta$ component in object space (spherical coordinates) into an *r* component (cylindrical coordinates) in the BFP[36], as depicted in Fig. 2(a).

To quantify the directionality of the antennas from the obtained BFP images, we computed the $F/B$ ratio. Different definitions can be used to calculate $F/B$ (see description in SI and comparison Fig. S1(a)), but in this paper we used the following one:

$$F/B = 10log_{10} \frac{\int_{\theta_1-\delta_1}^{\theta_1+\delta_1} \int_{\varphi_1-\delta_2}^{\varphi_1+\delta_2} S(\theta, \varphi) sin\theta d\theta d\varphi}{\int_{\theta_2-\delta_1}^{\theta_2+\delta_1} \int_{\varphi_2-\delta_2}^{\varphi_2+\delta_2} S(\theta, \varphi) sin\theta d\theta d\varphi} \quad \text{(dB)} \quad (1)$$

where $S(\theta,\varphi)$ represents the power radiated by the antenna in a given direction $(\theta,\varphi)$ per unit solid angle. Considering the distribution of the signal, we calculated the ratio of radiated power in two broad angular ranges ($(\theta_1 - \delta_1 \to \theta_1 + \delta_1, \varphi_1 - \delta_2 \to \varphi_1 + \delta_2)$ and $\theta_2 - \delta_1 \to \theta_2 + \delta_1, \varphi_2 - \delta_2 \to \varphi_2 + \delta_2)$) to quantify $F/B$ from eq. (1). Here, $(\theta_1, \varphi_1)$ corresponds to the angular position of the maximum lobe in the range $90° < \varphi < 270°$, whereas $(\theta_2, \varphi_2)$ is the direction of maximum signal in $\varphi \geq 270°$ or $\varphi \leq 90°$. If there is no lobe in that second angular region, then $\varphi_2 = \varphi_1 + \pi$. Considering the angular extent of the signal in the simulated BFP images, we chose $\delta_1$=10°, $\delta_2$=50°. The area enclosed within these values and used for the calculation of the $F/B$ ratio of antenna is marked with red sectors in Fig. 2(c, d). Using this definition, we computed the $F/B$ ratio as a function of wavelength for the NRDAs. Moreover, since directivity is a key factor in the description of directional antennas in radio wave applications[37], we also took this parameter into account (see comparison with $F/B$ values in Fig. S1(a)):

$$Dir_{max} = \frac{4\pi S_{max}(\theta, \varphi)}{\int_0^{2\pi}\int_0^{\pi} S(\theta, \varphi)sin\theta d\theta d\varphi} \quad (2)$$

where $Dir_{max}$ represents the ratio of maximum radiated power per unit solid angle $S_{max}(\theta, \varphi)$ to the average radiated power in $4\pi$ direction.

A comparison of these two parameters ($F/B$ and directivity) as a function of wavelength between NRMA and NRDA is shown in Fig. 2(b). For the case of the NRMA, the directivity is around $Dir_{max} \approx 7$ (or $Dir_{max} \approx 1.5$ in the absence of a substrate (Fig. S2(c)), as expected for an infinitesimal dipole antenna[37]) and the $F/B$ ratio is nearly 0 dB, showing no preferential emission direction. Conversely, for the NRDA, both $Dir_{max}$ and $F/B$ ratio show a peak at λ = 570nm. To explain the spectral position of this peak, we computed the scattering spectra of both monomer and dimer of NR with plane wave excitation (Fig. S2(a)). Since the transverse mode of the AuNRs is weaker than longitudinal one, and since both are quite close due to the small aspect ratio of AuNRs (1.7), only one scattering peak is observed. Due to mode hybridization of the dimer[38-40], the longitudinal plasmon band splits into two modes: one shifting to shorter wavelength forming the antibonding mode (bright mode or in-phase mode) and another one red-shifting as bonding mode (dark mode or antiphase mode)[41, 42]. The latter does not show up in the scattering spectra due to the side-by-side symmetry of the dimer when excited by a plane wave. However, in the case of the asymmetrical near-field excitation produced by locating one emitter at the tip of one AuNR, this symmetric condition is broken and the antiphase mode can be excited. Similar to the results from Pakizeh *et al.*[27] for a dimer of Au nanodisks system, we observe the maximum directivity near the antiphase mode. As expected, a peak in the $F/B$ ratio appears at λ = 570 nm, which is red-shifted compared to the longitudinal mode of a single nanorod and can then be attributed to the dimer's antiphase mode. We also calculated the radiation efficiency of both NRMA and NRDA, and found that NRDA showed lower radiation efficiency compared to NRMA (Fig. S2(b)). We associate this effect to higher ohmic losses in the assembly of two AuNRs. Considering the spectral dependence of both the $F/B$ ratio and the radiation efficiency, experimentally conditions might require detecting a range of frequencies which comes at the expense of decreasing the maximum obtainable directionality (Fig. S3).

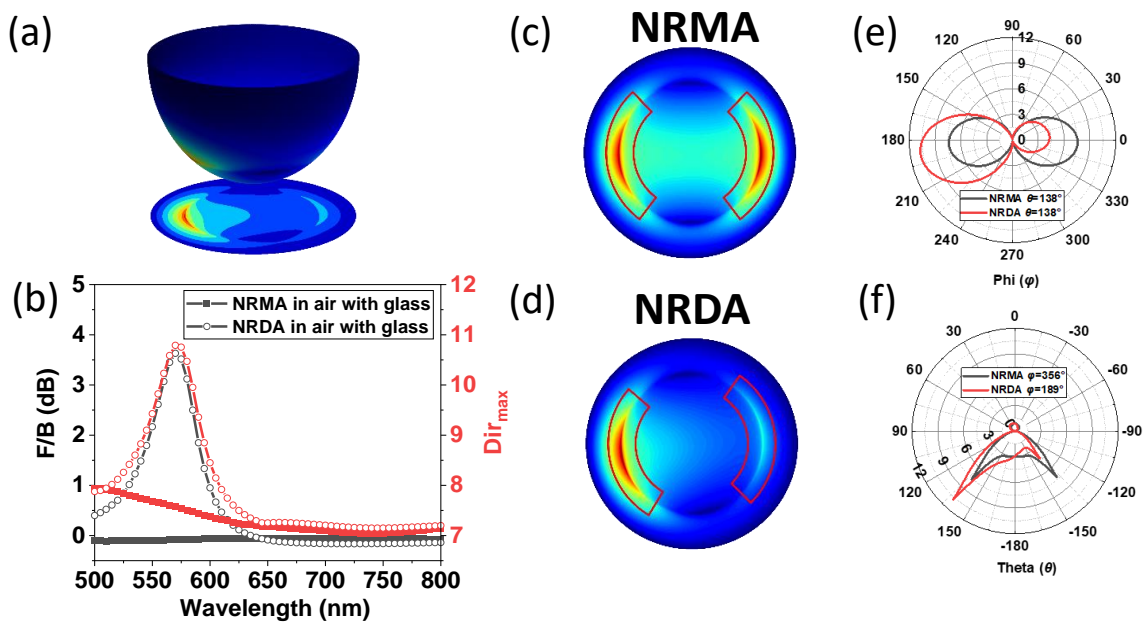

*Figure 2 (a) NRDA's radiation on a hemispherical surface and its projection on the back focal plane. (b) F/B ratio (black line) and maximum directivity (red line) of NRMA (solid square) and NRDA (hollow circle) in air with glass substrate. (c, d) Back focal plane images of NRMA and NRDA at 570 nm (wavelength of maximum directivity). (e, f) Corresponding polar radiation patterns of NRMA and NRDA with fixed theta (e) and phi (f) at 570 nm.*

Another way to visualize the radiation pattern of the ultracompact antennas is to use polar plots. Fig. 2(e) shows the azimuthal polar plot (φ = 0 to 360°) in the direction of maximum emission ($\theta$ = 138°) for both NRMA and NRDA. Conversely, Fig. 2(f) displays the altitudinal polar plot ($\theta$ = 0 to 180°) at the direction of maximum emission ($\varphi$ = 356° for NRMA and 189° for NRDA). Due to near-field interaction between the glass surface and the localized surface plasmon of the AuNRs, most of the evanescent field is radiated into the direction corresponding to the critical angle ($\theta_c$)[43-46] of air-glass interface, which is $\theta_c \approx 42°$.

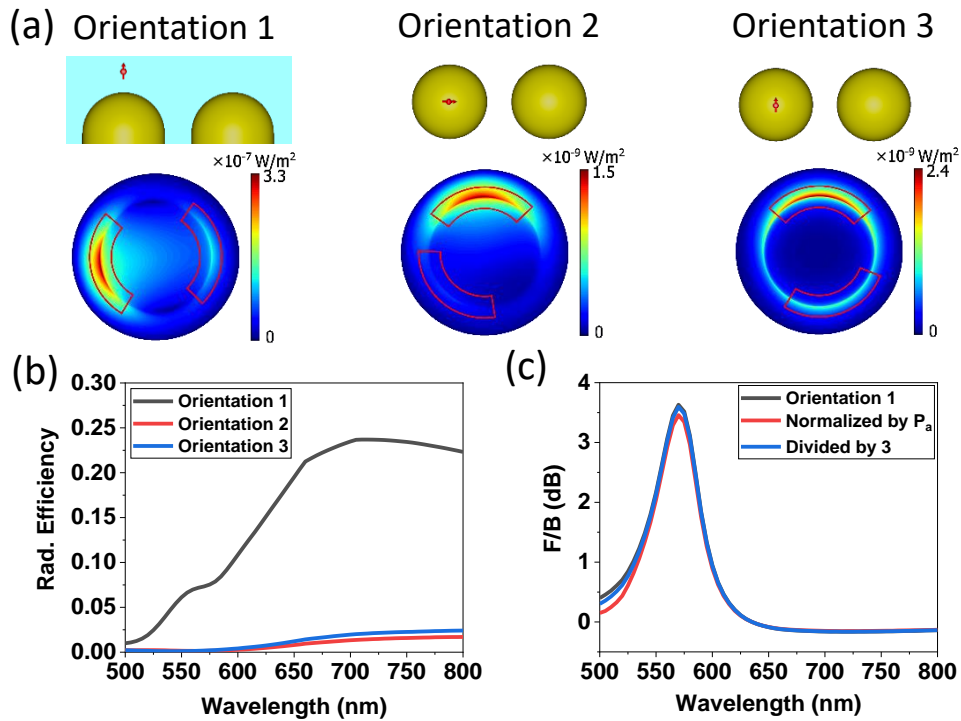

Figure 3 Effect of dipole orientation on NRDA's emission. (a) Schematics of dipole orientation (top) and corresponding back focal plane images at 570 nm (bottom). (b) Radiation efficiency of NRDA for different dipole orientations. (c) Spectral dependency of the F/B ratio for a dipole with "Orientation 1" and for an averaged rotated dipole, which is calculated using two different methods: normalizing the total radiated power by the accepted power ($P_a$) and then dividing by 3, or dividing the total radiative power by 3 directly.

In order to optimize unidirectionality and study how feasible it will be to achieve it in experimental conditions, we tuned several parameters for the dimer antenna. As it is well known, it is hard to controllably orient the dipole moment of nanoemitters such as fluorescent dyes. Thus, we first studied the effect on the directionality of the orientation of the emitter represented as a dipole. Models of NRDAs with three possible orthogonal orientations of the dipole are shown in Fig. 3(a), together with their corresponding BFP images at 570 nm (antiphase mode). By computing the radiation efficiency for each case, we observe that only when the emitter is oriented along the axial direction of the AuNR ("Orientation 1") it is not quenched (Fig. 3(b)). For the two remaining orientations, their radiation power contributed to the average radiation pattern accounts for less than 1%. Therefore, the average $F/B$ ratio detectable on the far-field is determined by "Orientation 1" (Fig. 3(c)).

Unlike single nanorods, dimers of NRs are more prone to exhibit deviations from the designed geometry under realistic fabrication conditions. This has an effect on the resonance properties too[40]. Therefore, we studied the influence of the different geometrical parameters: gap1 plays an important role in controlling the interaction between the AuNRs and the quantum emitter; gap2 controls the extent of the hybridization between both AuNRs; and gap3 determines the coupling between the antenna and the substrate. According to Fig. 4, fluctuations of gap1 (Fig. 4(a)) and gap3 (Fig. 4(c)) in a certain range (~ 10 nm) would not affect much the directionality of the NRDA, showing the robustness of the design. Conversely, reducing gap2 caused a stronger hybridization between the two AuNRs, which is manifested as a red-shift of the antiphase mode and as an improvement of the $F/B$ (Fig. 4(b)).

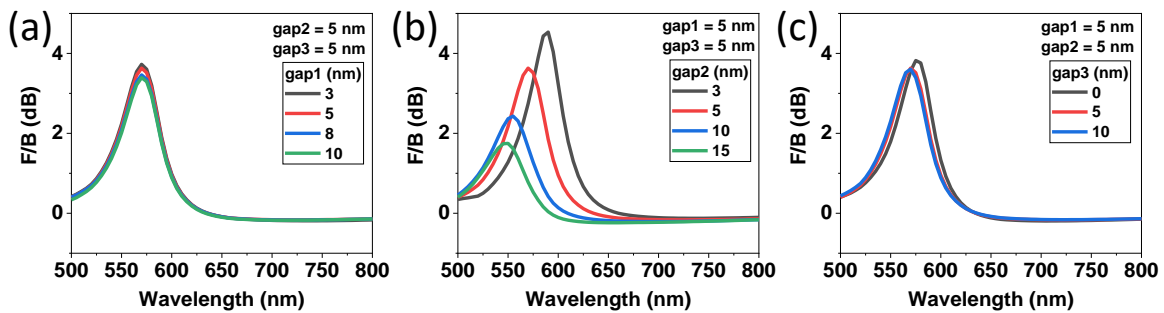

*Figure 4 Impact of the different NRDA gaps on the F/B ratio. (a) Effect of gap 1 (3, 5, 8, 10). (b) Effect of gap2 (3, 5, 10, 15). (c) Effect of gap3 (0, 5, 10).*

Nevertheless, unidirectionality is reduced but not fully lost even for the largest gap studied. We also noticed that tuning gap1 caused a significant non-radiative loss due to higher energy transfer and dissipation inside AuNRs, and that this is not affected by changes in gap2 and gap3 (Fig. S4).

      Other geometrical parameters that can experimentally change due to synthesis or fabrication are: translocation of one AuNR (Fig. 5(a)), in plane (x-y) movement of the emitter away from the tip centre (Fig. 5(b)), rotation of one AuNR (Fig. 5(c)) and size mismatch between both AuNRs (Fig. 5(d)). Despite these changes, some of these non-optimal dimers can still display emission directionality, as seen in Fig. 5. $F/B$ ratio changed only 0.5 dB when the second AuNR moved up along the y-direction (y from 0 to 30 nm, Fig. 5(a)). On the contrary, when this second AuNR is moved down along y-direction, the $F/B$ ratio can be 0 dB and even emission direction reverses, although the radiation efficiency of the antenna increased (Fig. S5(a)). On the other hand, $F/B$ values increased when the emitter is moved closer to the second AuNR along the x-direction, but were barely changed when the emitter was moved along the z-direction (Fig. 5(b)). According to Fig. 5(c), tilting of the AuNR has a noticeable effect in both the magnitude of the directionality as well as on the wavelength of the antiphase mode. Finally, enlarging the length of a single AuNR leads to a red-shift of the longitudinal mode, and so it does for the antiphase mode of the dimer[40]. Moreover, the $F/B$ ratio can get significantly increased in such case (Fig. 5(d)). The radiation efficiency at the wavelength of maximum directionality did not change much in this case (Fig. S5(d)), which is good to guarantee detection in experimental conditions. Interestingly, the radiation efficiency curve shows a dip, which is related to enhanced coupling strength and to a larger energy split between the in-phase and antiphase modes in the dimer. Besides, if only one of the AuNRs becomes longer, the maximum $F/B$ ratio changes only slightly (Fig. 5(d)).

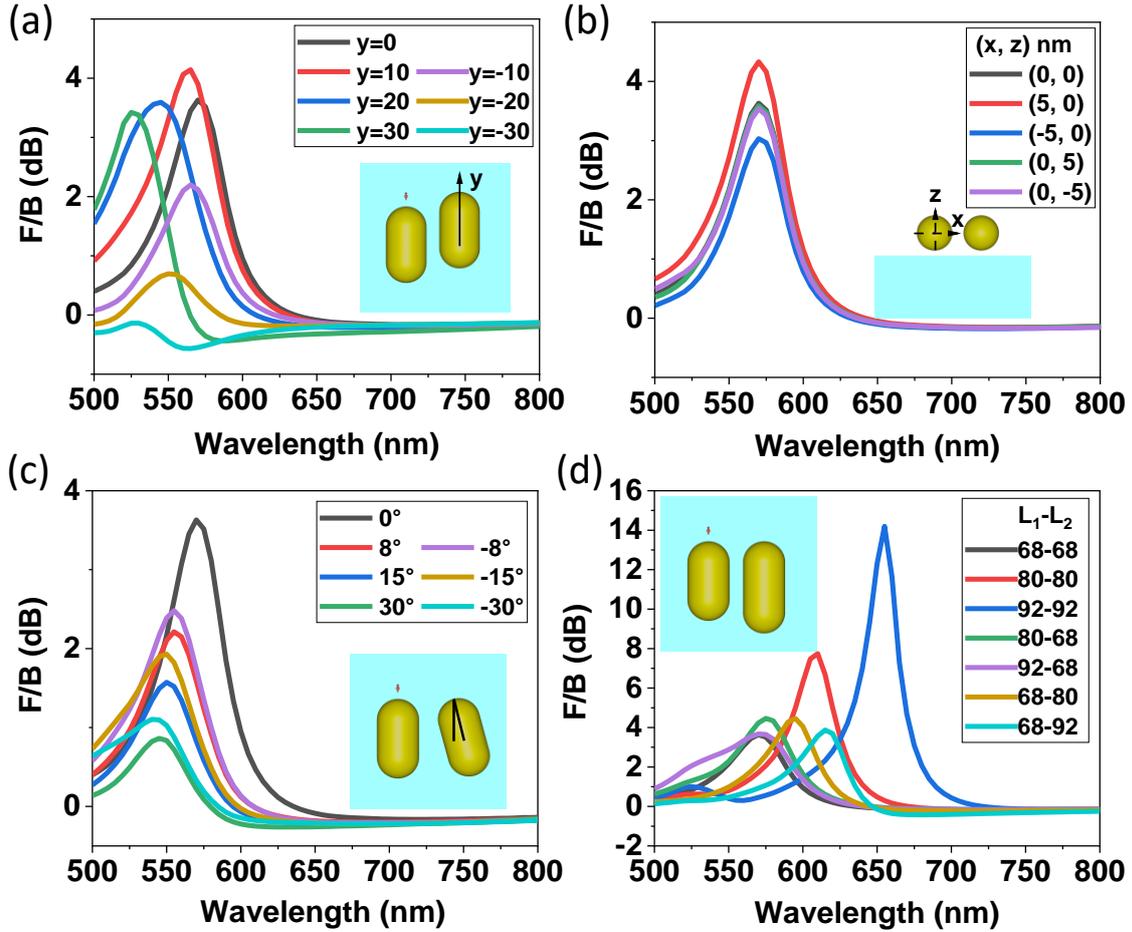

*Figure 5 Impact of different NRDAs configurations on F/B ratio. (a) Second nanorod moving along y direction (units are in nanometers). (b) Dipole moving on top of first nanorod away from its center (units are in nanometers). (c) Second nanorod rotating around its vertex. Minus symbol represents clockwise rotation. d) Combination of nanorods with different lengths. The two numbers in the legend represent the length of first and second nanorod, respectively (units are in nanometers).*

The last geometrical parameter that was studied was the tip curvature of AuNRs, which can produce different local electric fields and severely influences the nearby emitter[47]. We simulate the curvature of AuNR by adding some semi-spherical caps that have a radius of T = 20 nm. Then, we modify tip curvature by changing the length of the protrusion (T) and compressing this hemisphere caps to semi-ellipsoid while keeping the NR's total length (68 nm) constant. We found that not only the longitudinal mode shifted from 570 nm to 610 nm, but also unidirectionality changed from 3.6 dB to 7.5 dB as shown in Fig. S6(a). Contrary, the radiation efficiency at the wavelength of maximum directionality increased 13% (Fig. S6(b)).

Overall, these results show that the NRDA presented here, composed of two parallel nanorods and a dipole located at the tip of one of them, shows excellent unidirectional emission and is robust to fabrication variability and shape/size inhomogeneity of nanorods.

**Discussion**

Due to plasmon hybridization between both AuNRs, the phase delay produced in NRDA replaced the larger gap that is necessary in Yagi-Uda antennas to achieve constructive and destructive interference in the near-field (Fig. 6(a)). In order to explore the mechanism behind this phenomenon in further detail, we utilized a two dipole analytical model[20] to quantify this phase difference between both AuNRs. When energy is transferred from the dipole to the AuNRs in the near-field, photons are emitted through localized surface plasmons, which are collective oscillations of electrons induced on the surface of AuNRs. Here, we treated these surface plasmons as radiating electric dipoles. Due to the asymmetric configuration of the quantum emitter and the NRs dimer, each NR owned a different electric dipole moment. Thus, the overall system can be described by an amplitude change $\left(\frac{|P_1|}{|P_2|}\right)$, a phase delay coming from hybridization ($\Delta\varphi$), and a phase delay coming from the gap ($kd$, $k = \frac{2\pi n}{\lambda}$, with n being the refractive index of the surrounding medium). The gap here is the distance between the average centre of the dipole moments, and not the physical distance between the edges of both AuNRs. With these parameters, we quantified unidirectionality by computing the ratio between the dipole intensity at both sides (left and right) of the dimer:

$$\frac{I_L}{I_R} = \frac{|P_1 + e^{+ikd}P_2|^2}{|P_1 + e^{-ikd}P_2|^2} \qquad (3)$$

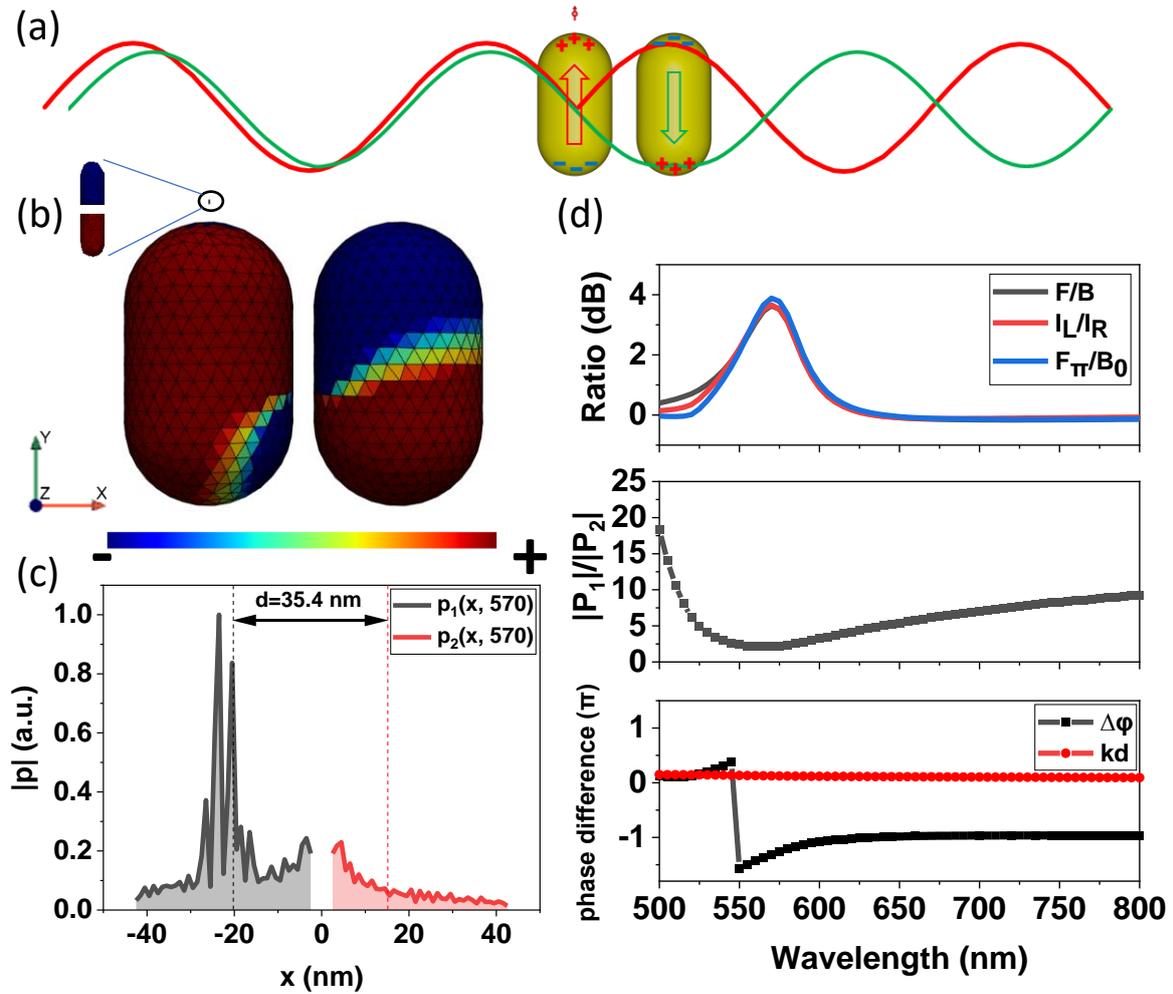

*Figure 6 Two dipole model. (a) Schematic representation of the unidirectional emission. Red and green lines represent the electric fields produced by the nanorod close to the dipole and by the second nanorod, respectively. Constructive and destructive interference happen on left and right sides, respectively. (b) Surface charge density distribution of model at 570 nm (inset shows the dipole emitter). (c) Dipole moment distributions of AuNRs along x direction at 570 nm. Black and red dash lines correspond to average center of dipole moment in first (left nanorod) and second (right nanorod) dipole. (d) F/B calculated by: simulation (black line), two dipole model (red line) and simulated intensity ratio (blue line) at $\varphi = \pi$ and $\varphi=0$ (top), ratio of the magnitude of total dipole moment in both nanorods (middle) and phase difference (bottom).*

According to the surface charge density distribution around the AuNRs (Fig. 6(b)), we found that the dipole moments in two AuNRs showed different orientations and amplitudes near 570 nm, in clear contrast to the symmetric excitation by a plane wave (Fig. S7(a)). Distribution at other wavelengths and for the NRMA cases are shown in Fig. S7. In NRDA, the gap between both dipole distributions was around 35.4 nm and less than the physical gap centre (40 nm) because of inhomogeneous distribution of charge on the AuNRs surface (Fig. 6(c)). The phase in second AuNR starts to reverse at a wavelength ($\lambda$ = 550 nm) that corresponds to the hybridized antiphase mode of the dimer. The total phase difference ($kd - \Delta\varphi$) is $1.43\pi$ and the amplitude change $\left(\frac{|P_1|}{|P_2|}\right)$ is 2.16, these two parameters together determined optimal directionality is at λ = 570 nm (Fig. 6(d)). These results from two dipole model $\left(\frac{I_L}{I_R}\right)$ are quite close to FEM simulated results ($F/B$ and $F_\pi/B_0$). Therefore, we can conclude that directionality of the dimer antenna stems from the antiphase mode under asymmetric excitation of a quantum emitter in the near-field.

As in the case of the FEM simulations, we can also study the influence of the different geometrical parameters on the behaviour of the system. When the AuNRs become longer (L = 92 nm, Fig. S8), the ratio between both dipole moments $\left(\frac{|P_1|}{|P_2|}\right)$ gets closer to 1 (1.06) and $(kd - \Delta\varphi) = 1.23\pi$. Hence, the antenna showed much higher unidirectionality at the resonance wavelength, in agreement with FEM simulations: $F/B$ goes from 3.6 dB (L = 68 nm) to 14.2 dB (L = 92 nm). To some extent, hybridization in a dimer of longer AuNRs was stronger and induced a higher energy transfer and more thorough destructive interference. Changes in other parameters such as the tip curvature, the glass substrate or the surrounding medium also showed good agreement between FEM simulations and the analytical two dipole model (Fig. S9-S12).

In conclusion, the proposed ultracompact NRDA shows higher $F/B$ when the ratio between both dipole moments $\left(\frac{|P_1|}{|P_2|}\right)$ is closer to 1 and $(kd - \Delta\varphi)$ is closer to $\pi$ at antiphase mode, which is also accompanied by a stronger mode hybridization in the dimer.

**Conclusion**

In summary, we have shown that ultracompact antennas based on two-parallel AuNRs display robust and excellent directionality within tolerable deviation from target configuration. Furthermore, the stronger the hybridization between both nanorods is, the higher the directionality of the antenna. The described dimer structure is easy to fabricate by various wet assembly methods[48] or AFM manipulation[30]. The most crucial component of this ultracompact NRDA is to precisely place the single emitter in the near-field of one of the nanorods, which is indispensable for asymmetric excitation of the anti-phase mode. This could be achieved for example via soft template assembly techniques, such as DNA origami[49, 50]. Overall, the ultracompact NRDA design provides a new possibility to further study antenna-assisted directional single-photon-sources for integrated photonic chips.

**Methods**

A frequency domain solver based on finite element method (FEM) in CST Studio Suite was used for the 3D full-wave simulation.

For the model without a substrate, the boundaries were set to open (add space) in the six faces. In presence of a substrate, the boundaries were set to open except for the plane wave input surface. The refractive index (n) of air, water and glass were set to n=1, 1.33 and 1.5, respectively. The dielectric function of gold was taken from fitting data of Johnson & Christy[51].

In the far-field simulations with substrate, the size of the glass substrate was 1000 x 1000 x 500 nm (length x width x thickness). A discrete port with 5000 ohms combined with Hertzian dipole were simulated as one single emitter.

For calculating the scattering spectra of AuNR with a glass substrate, the size of the substrate was changed to 400 x 400 x 150 nm. A material independent mesh group setting was used for the AuNRs and adaptive mesh refinement was turned off in order to keep the same mesh number. The whole model was used to calculate the total electric field and the magnetic field, and the model without AuNRs (keeping its shape but changing the material to same material as the background) was used to calculate the background electric and magnetic fields. After subtraction, the final scattering fields were used to calculate the scattered power and cross-section. The scattering spectra were averaged from the results obtained from using an excitation by two plane waves with orthogonal polarization sources at normal incidence.

According to Gauss' law, surface charge density can be obtained by $\rho = \varepsilon_0 \cdot (\boldsymbol{n} \cdot \boldsymbol{E}) = \varepsilon_0 \cdot (n_x \cdot E_x + n_y \cdot E_y + n_z \cdot E_z)$ [52]. Dipole moment $p_i(x, \lambda) = \iint_{NR_i(x, \lambda)} \rho(x, y, z, \lambda) y dy dz$ and total dipole moment $P_i(\lambda) = \int_{NR_i(\lambda)} p_i(x, \lambda) dx$ are complex values here[20]. Consequently, $\sum_x |p_i| \neq |P_i|$.

**References**

1.   Koenderink, A. F., Single-Photon Nanoantennas. *ACS Photonics* **2017,** *4* (4), 710-722.
2.   Kullock, R.; Ochs, M.; Grimm, P.; Emmerling, M.; Hecht, B., Electrically-driven Yagi-Uda antennas for light. *Nature Communications* **2020,** *11* (1), 115.
3.   Filter, R.; Slowik, K.; Straubel, J.; Lederer, F.; Rockstuhl, C., Nanoantennas for ultrabright single photon sources. *Opt Lett* **2014,** *39* (5), 1246-9.
4.   Singh, A.; de Roque, P. M.; Calbris, G.; Hugall, J. T.; van Hulst, N. F., Nanoscale Mapping and Control of Antenna-Coupling Strength for Bright Single Photon Sources. *Nano Lett* **2018,** *18* (4), 2538-2544.
5.   Lee, K. G.; Chen, X. W.; Eghlidi, H.; Kukura, P.; Lettow, R.; Renn, A.; Sandoghdar, V.; Götzinger, S., A planar dielectric antenna for directional single-photon emission and near-unity collection efficiency. *Nature Photonics* **2011,** *5* (3), 166-169.
6.   Novotny, L.; van Hulst, N., Antennas for light. *Nature Photonics* **2011,** *5* (2), 83-90.
7.   Sakat, E.; Wojszvzyk, L.; Greffet, J.-J.; Hugonin, J.-P.; Sauvan, C., Enhancing Light Absorption in a Nanovolume with a Nanoantenna: Theory and Figure of Merit. *ACS Photonics* **2020,** *7* (6), 1523-1528.
8.   Andersen, S. K. H.; Kumar, S.; Bozhevolnyi, S. I., Ultrabright Linearly Polarized Photon Generation from a Nitrogen Vacancy Center in a Nanocube Dimer Antenna. *Nano Letters* **2017,** *17* (6), 3889-3895.
9.   Baiyasi, R.; Goldwyn, H. J.; McCarthy, L. A.; West, C. A.; Hosseini Jebeli, S. A.; Masiello, D. J.; Link, S.; Landes, C. F., Coupled-Dipole Modeling and Experimental Characterization of Geometry-Dependent Trochoidal Dichroism in Nanorod Trimers. *ACS Photonics* **2021**.

10. Chen, W.; Roelli, P.; Hu, H.; Verlekar, S.; Amirtharaj, S. P.; Barreda, A. I.; Kippenberg, T. J.; Kovylina, M.; Verhagen, E.; Martinez, A.; Galland, C., Continuous-wave frequency upconversion with a molecular optomechanical nanocavity. *Science* **2021,** *374* (6572), 1264-1267.
11. Xomalis, A.; Zheng, X.; Chikkaraddy, R.; Koczor-Benda, Z.; Miele, E.; Rosta, E.; Vandenbosch, G. A. E.; Martinez, A.; Baumberg, J. J., Detecting mid-infrared light by molecular frequency upconversion in dual-wavelength nanoantennas. *Science* **2021,** *374* (6572), 1268-1271.
12. Saemisch, L.; Liebel, M.; van Hulst, N. F., Control of Vibronic Transition Rates by Resonant Single-Molecule-Nanoantenna Coupling. *Nano Lett* **2020,** *20* (6), 4537-4542.
13. Tanaka, Y. Y.; Kimura, T.; Shimura, T., Unidirectional emission of phase-controlled second harmonic generation from a plasmonic nanoantenna. *Nanophotonics* **2021,** *10* (18), 4601-4609.
14. Curto, A. G.; Volpe, G.; Taminiau, T. H.; Kreuzer, M. P.; Quidant, R.; van Hulst, N. F., Unidirectional Emission of a Quantum Dot Coupled to a Nanoantenna. *Science* **2010,** *329* (5994), 930-933.
15. Kosako, T.; Kadoya, Y.; Hofmann, H. F., Directional control of light by a nano-optical Yagi–Uda antenna. *Nature Photonics* **2010,** *4* (5), 312-315.
16. See, K. M.; Lin, F. C.; Chen, T. Y.; Huang, Y. X.; Huang, C. H.; Yesilyurt, A. T. M.; Huang, J. S., Photoluminescence-Driven Broadband Transmitting Directional Optical Nanoantennas. *Nano Lett* **2018,** *18* (9), 6002-6008.
17. Abedi, S.; Pakizeh, T., Packed Yagi-Uda nano-antennas using a unidirectional feed at visible wavelengths. *Opt. Lett.* **2017,** *42* (23), 4788-4791.
18. Shegai, T.; Johansson, P.; Langhammer, C.; Kall, M., Directional scattering and hydrogen sensing by bimetallic Pd-Au nanoantennas. *Nano Lett* **2012,** *12* (5), 2464-9.
19. Shegai, T.; Chen, S.; Miljkovic, V. D.; Zengin, G.; Johansson, P.; Kall, M., A bimetallic nanoantenna for directional colour routing. *Nat Commun* **2011,** *2*, 481.
20. Vercruysse, D.; Sonnefraud, Y.; Verellen, N.; Fuchs, F. B.; Di Martino, G.; Lagae, L.; Moshchalkov, V. V.; Maier, S. A.; Van Dorpe, P., Unidirectional Side Scattering of Light by a Single-Element Nanoantenna. *Nano Letters* **2013,** *13* (8), 3843-3849.
21. Vercruysse, D.; Zheng, X.; Sonnefraud, Y.; Verellen, N.; Di Martino, G.; Lagae, L.; Vandenbosch, G. A.; Moshchalkov, V. V.; Maier, S. A.; Van Dorpe, P., Directional fluorescence emission by individual V-antennas explained by mode expansion. *ACS Nano* **2014,** *8* (8), 8232-41.
22. Lu, G.; Wang, Y.; Chou, R. Y.; Shen, H.; He, Y.; Cheng, Y.; Gong, Q., Directional side scattering of light by a single plasmonic trimer. *Laser & Photonics Reviews* **2015,** *9* (5), 530-537.
23. Lai, Y. H.; Cui, X. M.; Li, N. N.; Shao, L.; Zhang, W.; Wang, J. F.; Lin, H. Q., Asymmetric Light Scattering on Heterodimers Made of Au Nanorods Vertically Standing on Au Nanodisks. *Advanced Optical Materials* **2021,** *9* (3).
24. Pezeshki, H.; Wright, A. J.; Larkins, E. C., Ultra‐compact and ultra‐broadband hybrid plasmonic‐photonic vertical coupler with high coupling efficiency, directivity, and polarisation extinction ratio. *IET Optoelectronics* **2021**, 1-9.
25. Taminiau, T. H.; Stefani, F. D.; van Hulst, N. F., Enhanced directional excitation and emission of single emitters by a nano-optical Yagi-Uda antenna. *Opt. Express* **2008,** *16* (14), 10858-10866.
26. Kasani, S.; Curtin, K.; Wu, N., A review of 2D and 3D plasmonic nanostructure array patterns: fabrication, light management and sensing applications. *Nanophotonics* **2019,** *8* (12), 2065-2089.
27. Pakizeh, T.; Käll, M., Unidirectional Ultracompact Optical Nanoantennas. *Nano Letters* **2009,** *9* (6), 2343-2349.
28. Liu, M.; Lee, T.-W.; Gray, S. K.; Guyot-Sionnest, P.; Pelton, M., Excitation of Dark Plasmons in Metal Nanoparticles by a Localized Emitter. *Physical Review Letters* **2009,** *102* (10), 107401.
29. Shen, H.; Lu, G.; He, Y.; Cheng, Y.; Gong, Q., Unidirectional enhanced spontaneous emission with metallo-dielectric optical antenna. *Optics Communications* **2017,** *395*, 133-138.
30. Shen, H.; Chou, R. Y.; Hui, Y. Y.; He, Y.; Cheng, Y.; Chang, H.-C.; Tong, L.; Gong, Q.; Lu, G., Directional fluorescence emission from a compact plasmonic-diamond hybrid nanostructure. *Laser & Photonics Reviews* **2016,** *10* (4), 647-655.

31. Bonod, N.; Devilez, A.; Rolly, B.; Bidault, S.; Stout, B., Ultracompact and unidirectional metallic antennas. *Physical Review B* **2010,** *82* (11).
32. Zhang, T.; Xu, J.; Deng, Z. L.; Hu, D.; Qin, F.; Li, X., Unidirectional Enhanced Dipolar Emission with an Individual Dielectric Nanoantenna. *Nanomaterials (Basel)* **2019,** *9* (4).
33. Alaee, R.; Albooyeh, M.; Tretyakov, S.; Rockstuhl, C., Phase-change material-based nanoantennas with tunable radiation patterns. *Opt Lett* **2016,** *41* (17), 4099-102.
34. Yao, K.; Liu, Y., Controlling Electric and Magnetic Resonances for Ultracompact Nanoantennas with Tunable Directionality. *ACS Photonics* **2016,** *3* (6), 953-963.
35. Kerker, M.; Wang, D. S.; Giles, C. L., Electromagnetic scattering by magnetic spheres. *J. Opt. Soc. Am.* **1983,** *73* (6), 765-767.
36. Lieb, M. A.; Zavislan, J. M.; Novotny, L., Single-molecule orientations determined by direct emission pattern imaging. *J. Opt. Soc. Am. B* **2004,** *21* (6), 1210-1215.
37. Balanis, C. A., *Antenna theory: analysis and design*. John wiley & sons: 2015.
38. Prodan, E.; Radloff, C.; Halas, N. J.; Nordlander, P., A hybridization model for the plasmon response of complex nanostructures. *Science* **2003,** *302* (5644), 419-22.
39. Jain, P. K.; Eustis, S.; El-Sayed, M. A., Plasmon Coupling in Nanorod Assemblies: Optical Absorption, Discrete Dipole Approximation Simulation, and Exciton-Coupling Model. *The Journal of Physical Chemistry B* **2006,** *110* (37), 18243-18253.
40. Basyooni, M. A.; Ahmed, A. M.; Shaban, M., Plasmonic hybridization between two metallic nanorods. *Optik* **2018,** *172*, 1069-1078.
41. Li, J. N.; Liu, T. Z.; Zheng, H. R.; Gao, F.; Dong, J.; Zhang, Z. L.; Zhang, Z. Y., Plasmon resonances and strong electric field enhancements in side-by-side tangent nanospheroid homodimers. *Opt Express* **2013,** *21* (14), 17176-85.
42. Flauraud, V.; Bernasconi, G. D.; Butet, J.; Alexander, D. T. L.; Martin, O. J. F.; Brugger, J., Mode Coupling in Plasmonic Heterodimers Probed with Electron Energy Loss Spectroscopy. *ACS Nano* **2017,** *11* (4), 3485-3495.
43. Lukosz, W.; Kunz, R. E., Light emission by magnetic and electric dipoles close to a plane interface. I. Total radiated power. *J. Opt. Soc. Am.* **1977,** *67* (12), 1607-1615.
44. Lukosz, W.; Kunz, R. E., Light emission by magnetic and electric dipoles close to a plane dielectric interface. II. Radiation patterns of perpendicular oriented dipoles. *J. Opt. Soc. Am.* **1977,** *67* (12), 1615-1619.
45. Lukosz, W., Light emission by magnetic and electric dipoles close to a plane dielectric interface. III. Radiation patterns of dipoles with arbitrary orientation. *J. Opt. Soc. Am.* **1979,** *69* (11), 1495-1503.
46. Hellen, E. H.; Axelrod, D., Fluorescence Emission at Dielectric and Metal-Film Interfaces. *J Opt Soc Am B* **1987,** *4* (3), 337-350.
47. Kern, A. M.; Martin, O. J., Excitation and reemission of molecules near realistic plasmonic nanostructures. *Nano Lett* **2011,** *11* (2), 482-7.
48. Hou, S.; Zhang, H.; Yan, J.; Ji, Y.; Wen, T.; Liu, W.; Hu, Z.; Wu, X., Plasmonic circular dichroism in side-by-side oligomers of gold nanorods: the influence of chiral molecule location and interparticle distance. *Phys Chem Chem Phys* **2015,** *17* (12), 8187-93.
49. Kuzyk, A.; Jungmann, R.; Acuna, G. P.; Liu, N., DNA Origami Route for Nanophotonics. *ACS Photonics* **2018,** *5* (4), 1151-1163.
50. Rothemund, P. W., Folding DNA to create nanoscale shapes and patterns. *Nature* **2006**, *440* (7082), 297-302.
51. Johnson, P. B.; Christy, R. W., Optical Constants of the Noble Metals. *Physical Review B* **1972**, *6* (12), 4370-4379.
52. Huang, Y.; Ringe, E.; Hou, M.; Ma, L.; Zhang, Z., Near-field mapping of three-dimensional surface charge poles for hybridized plasmon modes. *AIP ADVANCES* **2015**, *5* (10), 107221.

# Supplementary Information

**Optical ultracompact directional antenna based on a dimer nanorod structure**

*Fangjia Zhu[1*], María Sanz-Paz[1], Antonio Fernández-Domínguez[2], Mauricio Pilo-Pais[1] and Guillermo P. Acuna[1*]*

[1] Department of Physics, University of Fribourg, Chemin du Musée 3, Fribourg CH-1700, Switzerland.

[2] Departamento de Física Teórica de la Materia Condensada and Condensed Matter Physics Center (IFIMAC), Universidad Autónoma de Madrid, E-28049 Madrid, Spain.

**Methods for calculating the $F/B$ ratio**

To quantify the unidirectionality of the antennas from the obtained BFP images, we computed the $F/B$ ratio using different definitions:

$$F/B = 10 log_{10} \frac{\int_{\theta_1-\delta_1}^{\theta_1+\delta_1} \int_{\varphi_1-\delta_2}^{\varphi_1+\delta_2} S(\theta, \varphi) sin\theta d\theta d\varphi}{\int_{\theta_2-\delta_1}^{\theta_2+\delta_1} \int_{\varphi_2-\delta_2}^{\varphi_2+\delta_2} S(\theta, \varphi) sin\theta d\theta d\varphi} \quad (dB) \quad (1)$$

$$F_\pi/B_0 = 10 log_{10} \frac{S(\theta_1, \pi)}{S(\theta_1, 0)} \quad (dB) \quad (2)$$

$$F_p/B_p = 10 log_{10} \frac{S(\theta_1, \varphi_1)}{S(\theta_1, \varphi_1-\pi)} \quad (dB) \quad (3)$$

$$F_1/B_2 = 10 log_{10} \frac{S(\theta_1, \varphi_1)}{S(\theta_2, \varphi_2)} \quad (dB) \quad (4)$$

$$(5)$$

$S(\theta, \varphi)$ represents radiated power of the antenna in a given direction $(\theta, \varphi)$ per unit solid angle. Considering that the signal collected comes from a broad angular range, the ratio of radiated power in two angular ranges (($\theta_1 - \delta_1 \rightarrow \theta_1 + \delta_1, \varphi_1 - \delta_2 \rightarrow \varphi_1 + \delta_2$) and $\theta_2 - \delta_1 \rightarrow \theta_2 + \delta_1, \varphi_2 - \delta_2 \rightarrow \varphi_2 + \delta_2$)) were calculated to quantify forward to backward ratio ($F/B$) from eq. (1). ($\theta_1, \varphi_1$) corresponds to angular position of the maximum lobe in the range $90° < \varphi < 270°$, whereas ($\theta_2, \varphi_2$) is the direction of maximum signal in $\varphi \geq 270°$ or $\varphi \leq 90°$. If there is no lobe in that second angular region, then $\varphi_2 = \varphi_1 + \pi$. Considering the angular extent of the signal in the simulated BFP images, we chose $\delta_1=10°$, $\delta_2=50°$.

Similarly, eq. (2) describes the radiated power ratio between the direction with $\varphi_1 = \pi$ and inverse the direction with $\varphi_2 = 0$, while eq. (3) computes the radiated power ratio between the direction with maximum radiated power and the inverse direction with same $\theta$. Finally, eq. (4) outputs the ratio between the radiated power in the direction of maximum lobe in two half spaces.

We computed the $F/B$ ratio as a function of wavelength for antennas using the above-described equations. All the obtained values were put together in Fig. S1(a). It is worth noticing that regardless of the method used for the $F/B$ quantification, the maximum value always occurred at the same wavelength (λ=570nm). Furthermore, the maximum point in the two lobes were not always in opposite directions, as shown in Fig. S1(b). For some cases, one point might not be enough to represent the whole intensity on one side. For those reasons, we mainly used F/B ratio in eq. (1) to quantify unidirectionality of antenna.

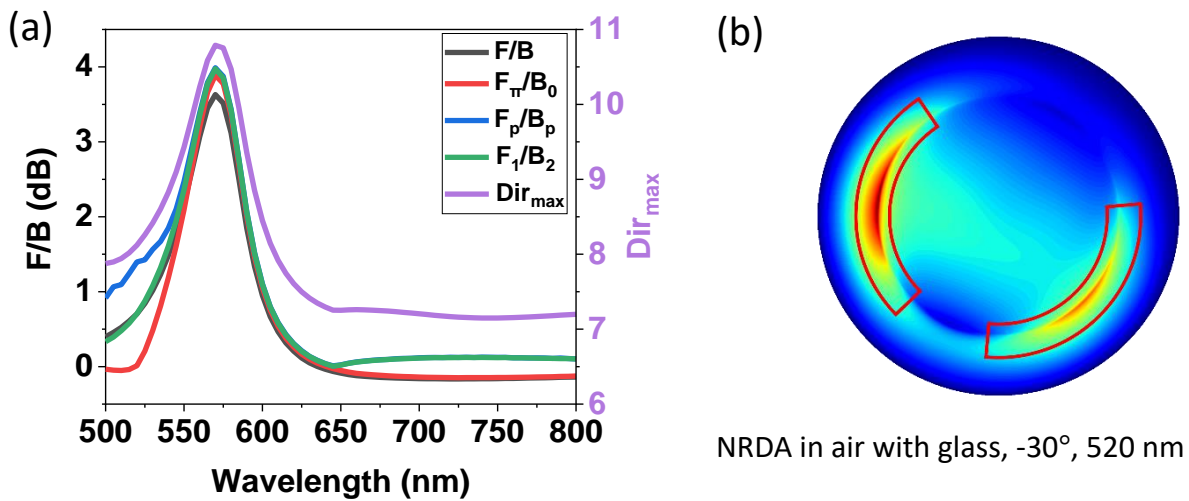

NRDA in air with glass, -30°, 520 nm

*Figure S1 (a) Comparison of different quantified methods. (b) Back focal plane images of dimer with first nanorod rotating clockwise 30°.*

**Optical properties of monomers and dimers with and without glass substrate**

We computed the scattering spectra of both monomer and dimer of nanorod with and without the glass substrate (Fig. S2(a)). The absence of glass substrate only caused a slight blue-shift for both monomer and dimer.

Fig. S2(b) describes the radiation efficiency of NRMA and NRDA with and without glass substrate. We also calculated the $F/B$ values for NRMA and NRDA in the absence of substrate (Fig. S2(c)). Fluctuations of the $F/B$ curve in air are due to the low signal collected on the second lobe. This does not happen for the case with glass substrate, where most of the energy is concentrated in a narrow range, causing a smoothening of the $F/B$ curve and an increased directivity.

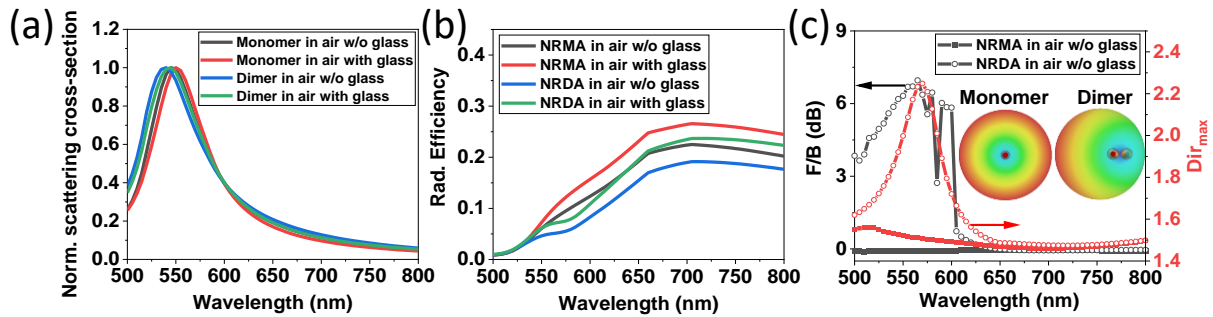

*Figure S2 Optical properties of monomer and dimer antenna in air in the presence or absence of a glass substrate. (a) Normalized scattering spectra of monomer and dimer. (b) Radiation efficiency of NRMA and NRDA. (c) F/B ratio (black line) and maximum directivity (red line) of NRMA (solid square) and NRDA (hollow circle) in air without glass substrate. Inserted images correspond to top view radiation patterns.*

**Changes in measured $F/B$ ratio depending on the wavelength range selected**

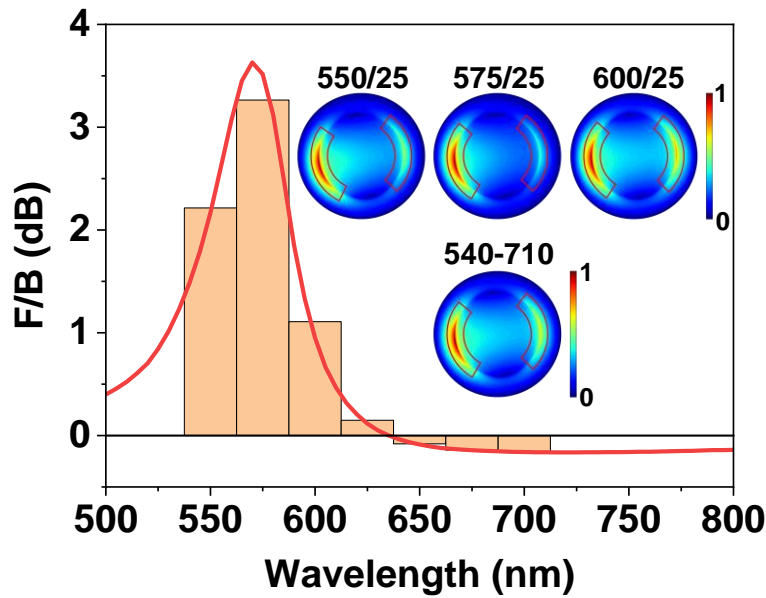

*Figure S3 Back focal plane images with bandpass filter. Red line represents F/B ratio of NRDA at every wavelength and histogram represents the F/B ratio within 25nm bandpass filter. Inserts are back focal plane image of 550/25, 575/25, 600/25 and broad range measurement without filter.*

In practice, since the antennas show unidirectionality only within a certain wavelength range, the common approach would be to collect the signal through a bandpass filter to reduce background and increase image contrast. Since the bandwidth of maximum $F/B$ ratio is narrow and the radiation efficiency depends on the wavelength, the measured unidirectionality will be significant different depending on the wavelength range considered. Thus, taking the spectral differences in radiation efficiencies into account, we calculated $F/B$ ratio within some bandpass filters that have a 25 nm bandwidth (Fig. S3). The value obtained for the bandpass filter containing the antiphase mode (550/25) showed only a slight difference from single wavelength (570 nm) results. However, if no bandpass filter is used, the final directionality gets reduced by nearly half: $F/B$ ratio goes from 3.3 dB in the case of the bandpass filter located at the antiphase mode, to 1.5 dB when collecting all spectral signal (540 -710 nm).

## Radiation efficiency of NRDAs as a function of gap1, gap2 and gap3

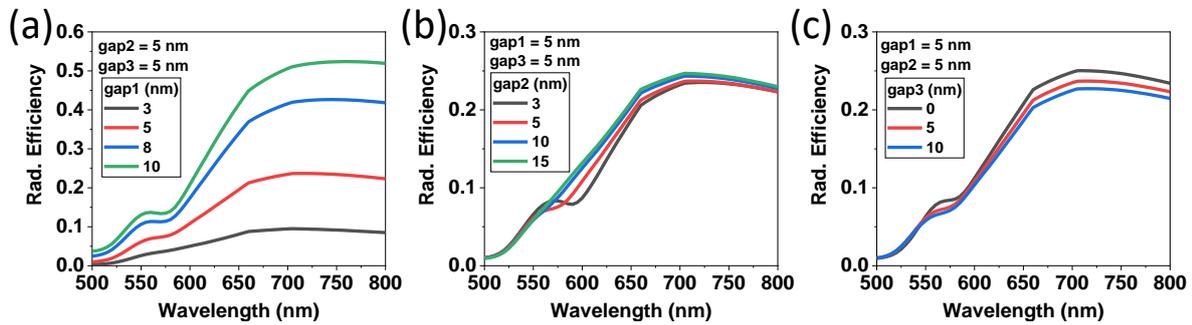

Figure S4 NRDAs with different gaps' impacts on radiation efficiency. Three number represent gap1, gap2 and gap3 in order. (a) Effect of gap 1 (3, 5, 8, 10). (b) Effect of gap2 (3, 5, 10, 15). (c) Effect of gap3 (0, 5, 10).

## Radiation efficiency of NRDAs as a function configuration

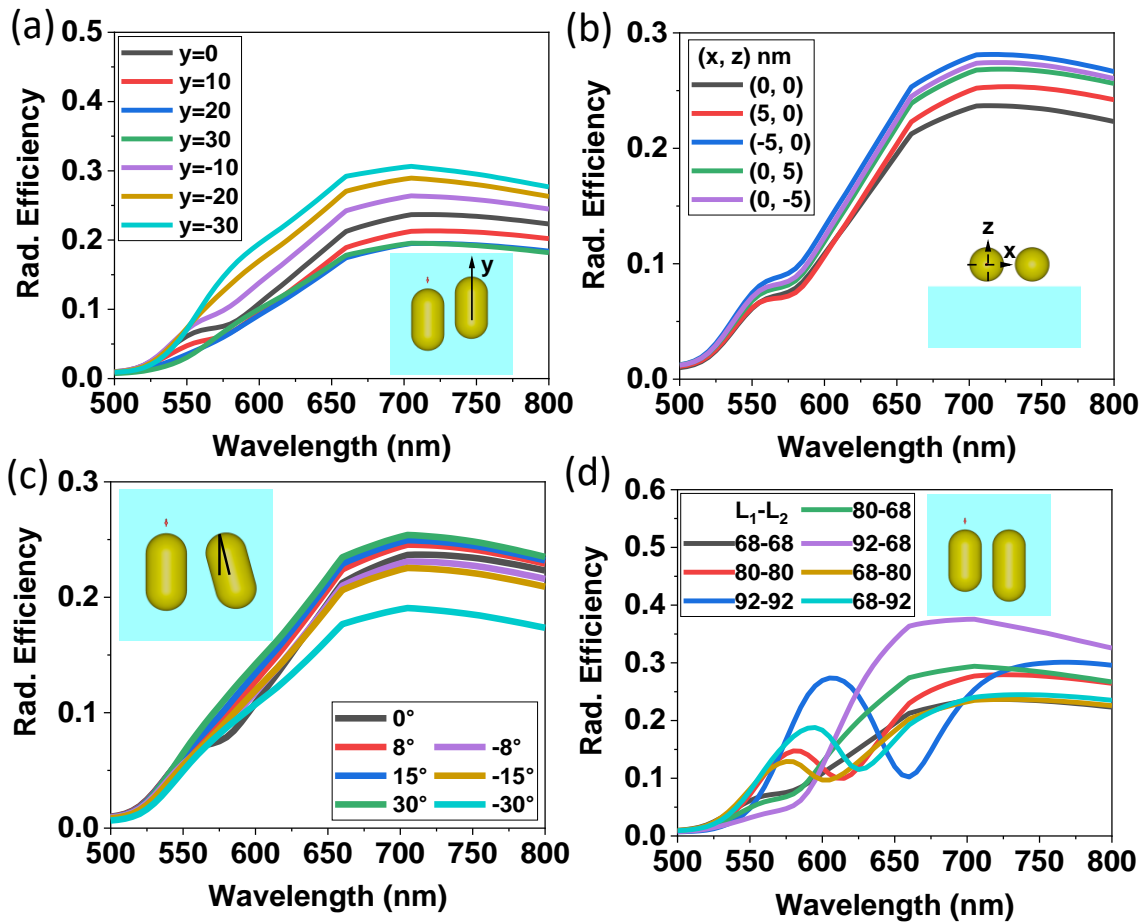

Figure S5 NRDA with different configurations' impacts on radiation efficiency. (a) Second nanorod move along y direction. Unit is nanometer. (b) Dipole moves on the top of nanorod. Unit is nanometer. (c) Rotation of nanorod. Minu symbol represents first nanorod rotates clockwise. Rotation origin is set as vertex of nanorod. (d) Combination of nanorods with different length. Two number represent length of first and second nanorod in order.

## Effect of tip curvature of AuNRs on NRDA's properties

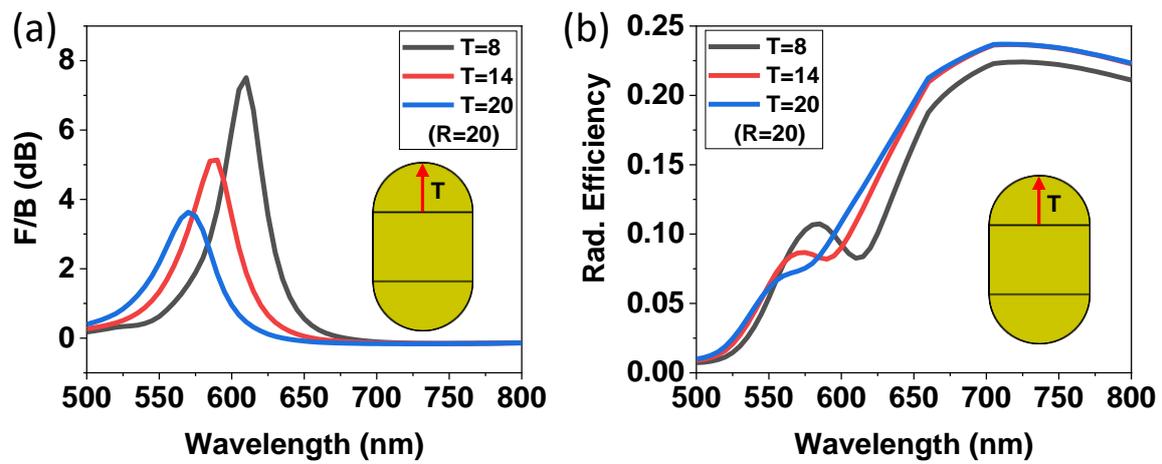

*Figure S6 Effect of nanorod's tip curvature in directionality (a) and radiated efficiency (b). Inserts represent cap of nanorod. Keep total length of nanorod as constant, compressing hemisphere to semi-ellipsoid to control the length of protruding tip (T).*

**Surface charge density distribution of excited two nanorods by plane wave and surface charge density distribution of NRDA and NRMA.**

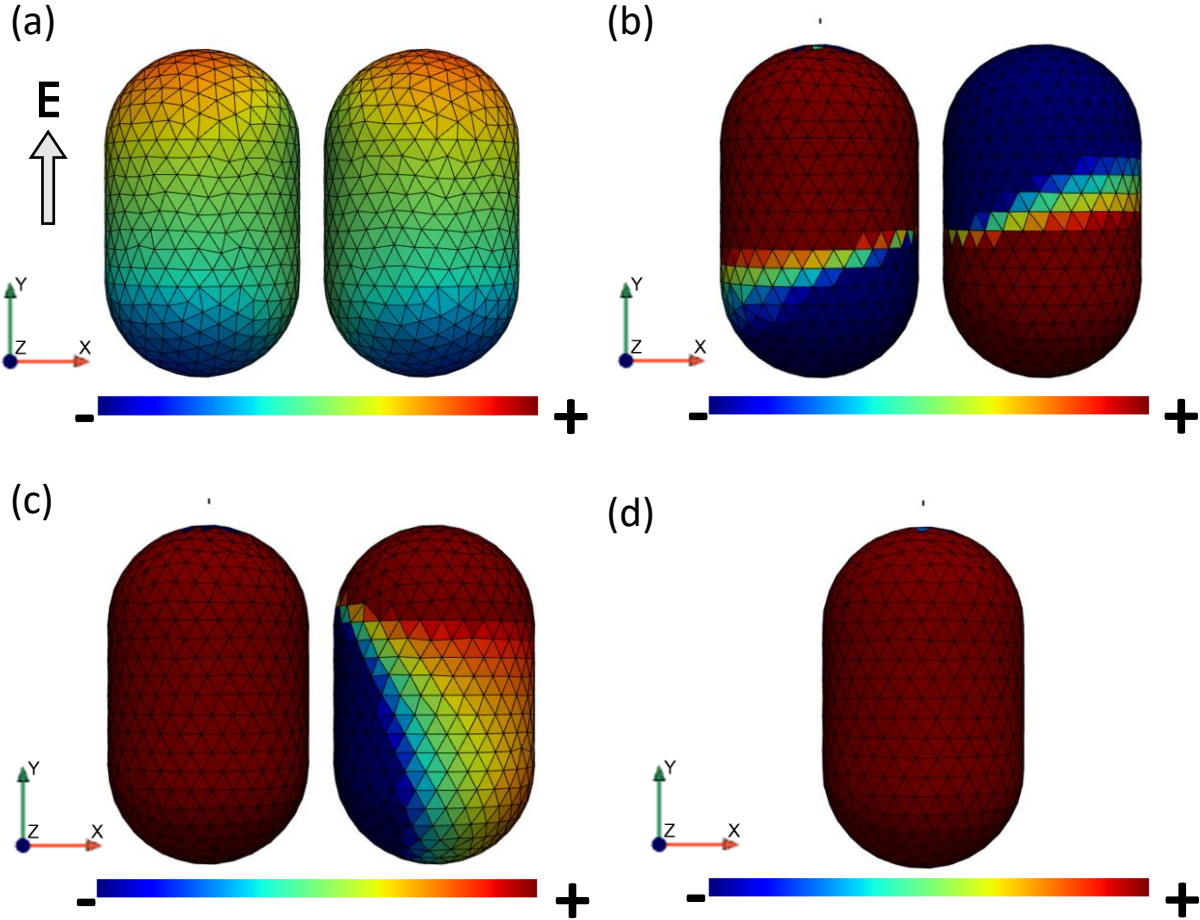

*Figure S7 (a) Surface charge density distribution of two nanorods (dimer) excited by 570 nm plane wave in air without glass substrate. Surface charge density distribution of NRDA at (b) 550 nm, (c) 700 nm and of NRMA at (d) 570 nm with glass substrate in air.*

**Effect of length of AuNR, glass substrate, tip curvature and surrounding medium in the two-dipole model results**

Surface charge density distribution of longer (L = 92nm) NRDA and corresponding analysis of dipole moment inside nanorods (Fig. S8). Parameters ($\frac{|P_1|}{|P_2|}$, $kd$, $\Delta\varphi$) and directionality didn't show significant difference for antenna without glass substrate (Fig. S9). In other words, the glass substrate didn't participate into antenna's directional emission directly.

When the curvature radius of AuNR is increased (T= 8, Fig. S10), the resonance wavelength of the antenna is red-shifted even with the same aspect ratio of the AuNRs. Besides, the antiphase mode is also red-shifted and the $F/B$ ratio enhanced to 7.5 dB ($F_\pi/B_0$ changed to 8.3 dB) compared to AuNR with the hemisphere tip. As in the previous case, smaller $\frac{|P_1|}{|P_2|}$ (1.32) caused higher unidirectionality.

When the antenna was immersed in water (Fig. S11), the resonance wavelength of AuNR red-shifted due to increased dielectric environment between gap[1, 2] as expected. Meantime, the antenna showed higher unidirectionality ($F/B$=14.8 dB, $F_\pi/B_0$ = 19.5 dB) due to smaller $\frac{|P_1|}{|P_2|}$ (1.07). In this case, unidirectionality and working wavelength can be manipulated by surrounding environment.

Similar analysis of NRDA with smaller gap2 are in Fig. S12.

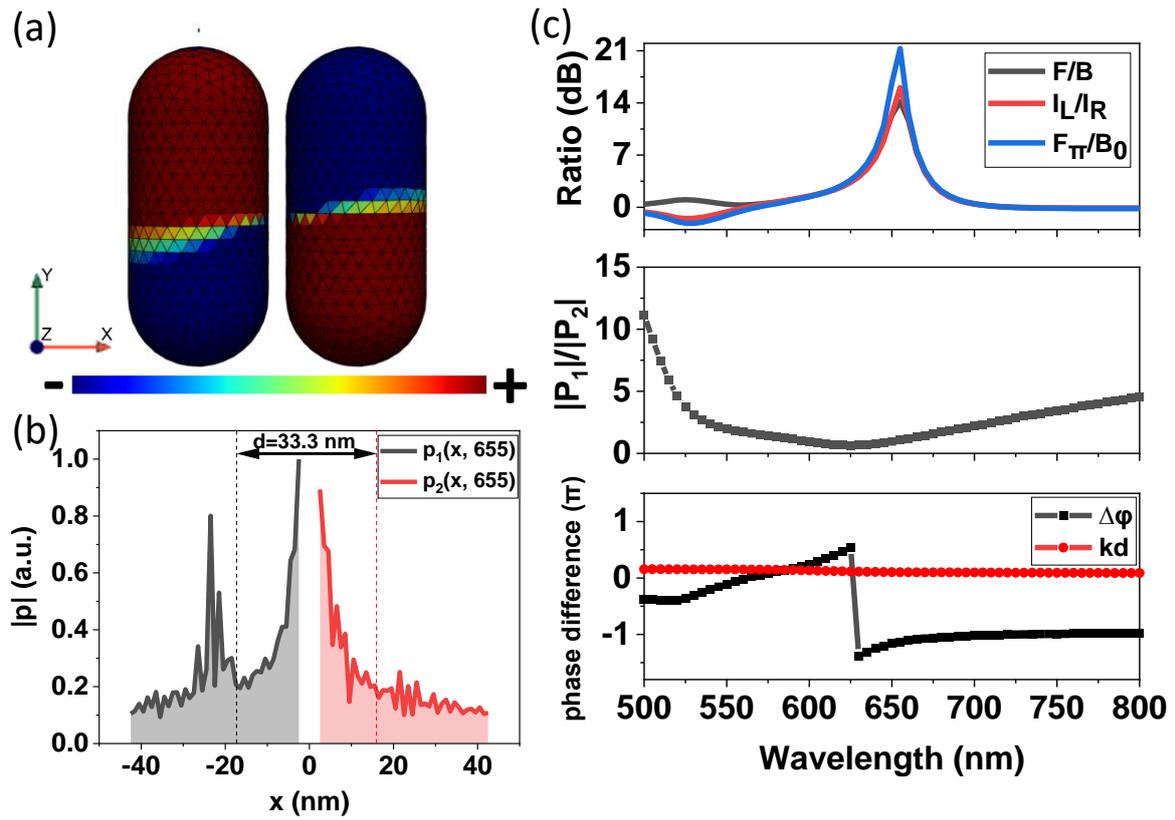

*Figure S8 Two dipole model. (a) Surface charge density distribution of model at 655 nm (with glass substrate, L=92 nm). (b) Dipole moment distributions of AuNRs along x direction at 655 nm. Black and red dash lines correspond to average center of dipole moment in first (left nanorod) and second (right nanorod) dipole. (c) F/B calculated by: simulation (black line), two dipole model (red line) and simulated intensity ratio (blue line) at $\varphi = \pi$ and $\varphi=0$ (top), ratio of the magnitude of total dipole moment in both nanorods (middle) and phase difference (bottom).*

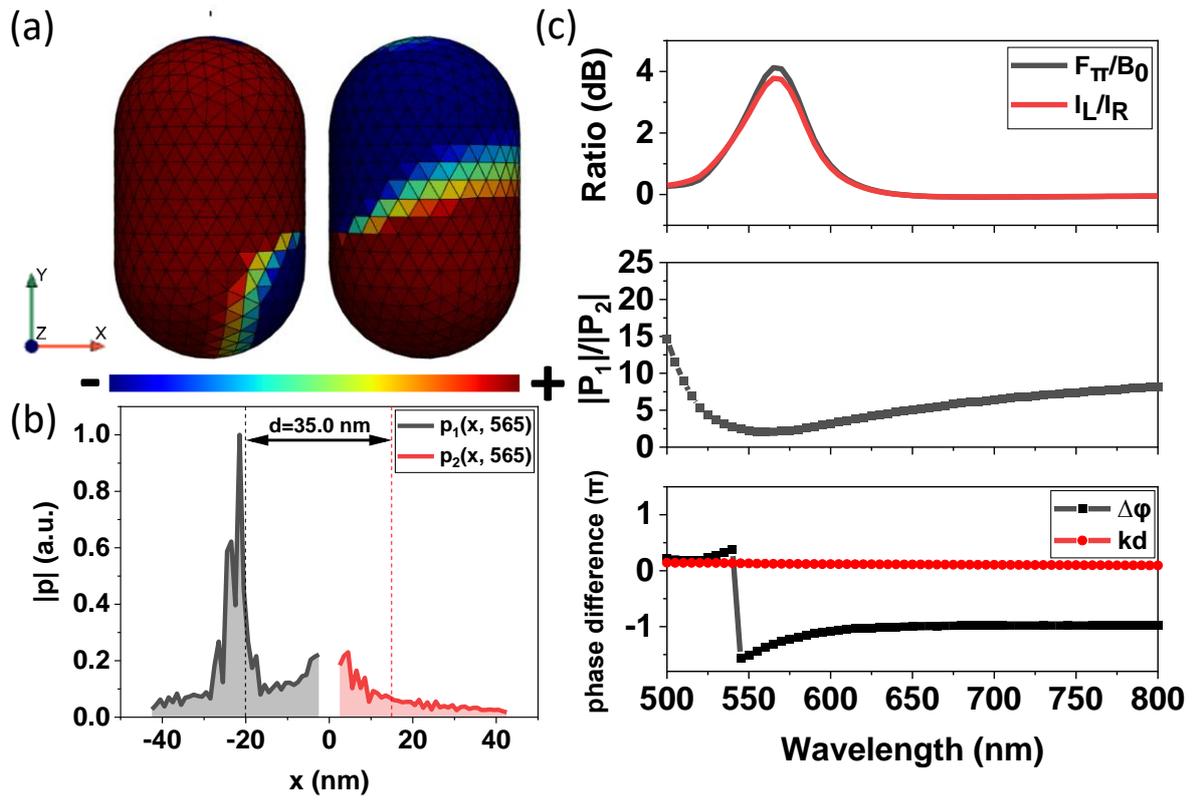

*Figure S9 Two dipole model. (a) Surface charge density distribution of model at 565 nm (without glass substrate in air). (b) Dipole moment distributions of AuNRs along x direction at 565 nm. Black and red dash lines correspond to average center of dipole moment in first (left nanorod) and second (right nanorod) dipole. (c) F/B calculated by: simulated intensity ratio (black line) at $\varphi = \pi$ and $\varphi=0$ (black line) and two dipole model (red line) (top), ratio of the magnitude of total dipole moment in both nanorods (middle) and phase difference (bottom).*

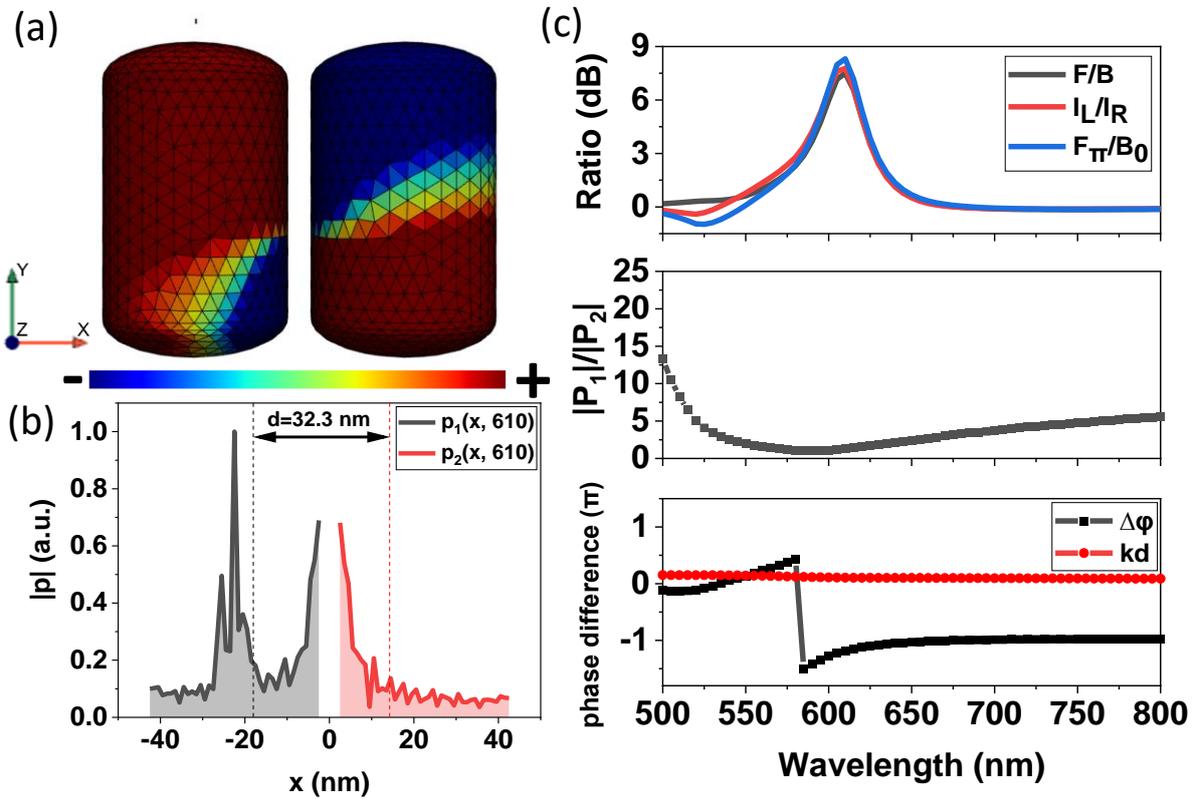

*Figure S10 Two dipole model. (a) Surface charge density distribution of model at 610 nm (with glass substrate, T=8 nm). (b) Dipole moment distributions of AuNRs along x direction at 610 nm. Black and red dash lines correspond to average center of dipole moment in first (left nanorod) and second (right nanorod) dipole. (c) F/B calculated by: simulation (black line), two dipole model (red line) and simulated intensity ratio (blue line) at φ = π and φ=0 (top), ratio of the magnitude of total dipole moment in both nanorods (middle) and phase difference (bottom).*

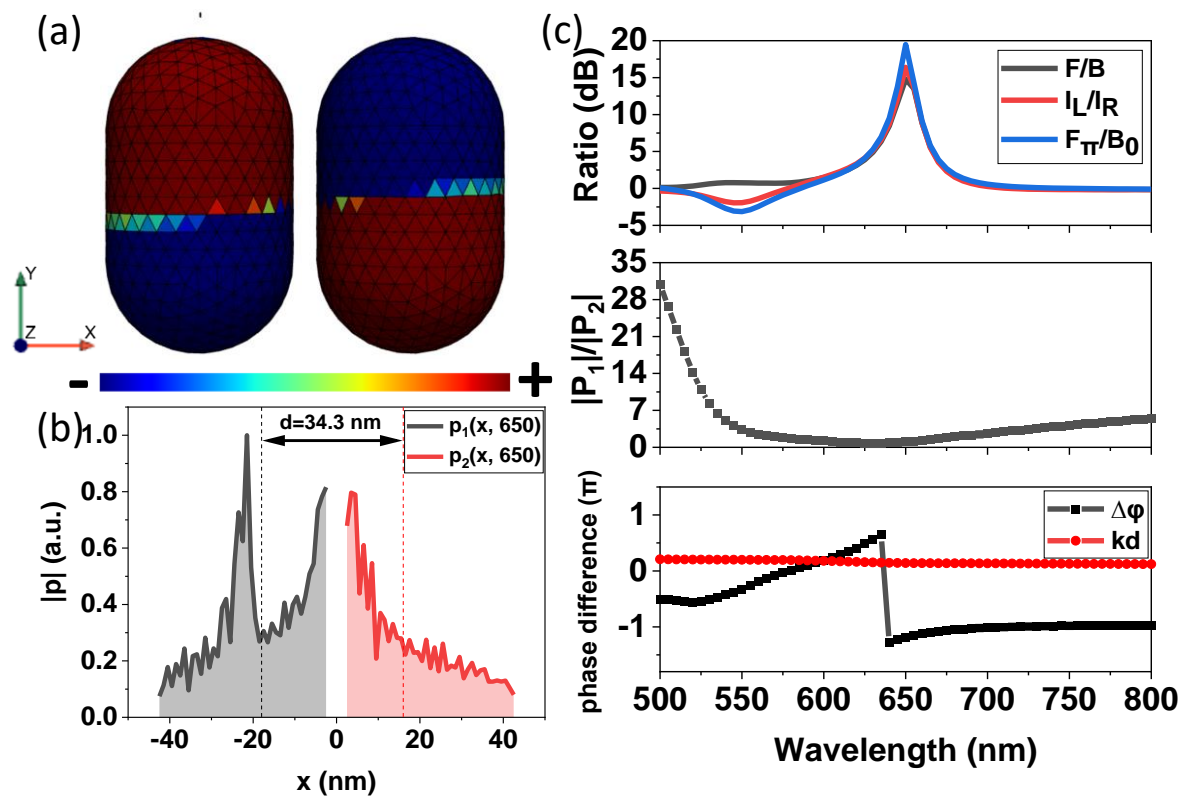

*Figure S11 Two dipole model. (a) Surface charge density distribution of model at 650 nm (with glass substrate, in water). (b) Dipole moment distributions of AuNRs along x direction at 650 nm. Black and red dash lines correspond to average center of dipole moment in first (left nanorod) and second (right nanorod) dipole. (c) F/B calculated by: simulation (black line), two dipole model (red line) and simulated intensity ratio (blue line) at φ = π and φ=0 (top), ratio of the magnitude of total dipole moment in both nanorods (middle) and phase difference (bottom).*

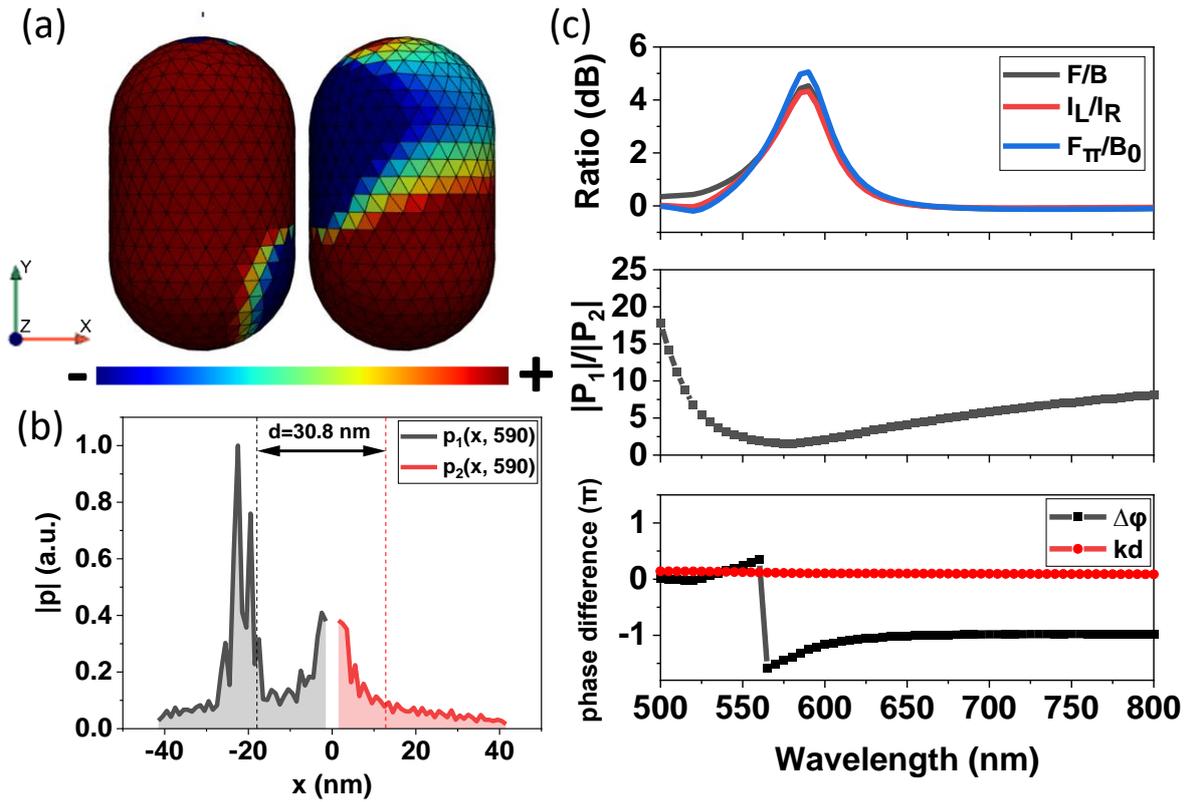

*Figure S12 Two dipole model. (a) Surface charge density distribution of model at 590 nm (with glass substrate, in air, gap2 = 3). (b) Dipole moment distributions of AuNRs along x direction at 590 nm. Black and red dash lines correspond to average center of dipole moment in first (left nanorod) and second (right nanorod) dipole. (c) F/B calculated by: simulation (black line), two dipole model (red line) and simulated intensity ratio (blue line) at $\varphi = \pi$ and $\varphi=0$ (top), ratio of the magnitude of total dipole moment in both nanorods (middle) and phase difference (bottom).*

## Radiated power enhancement in the presence of the antenna

Apart from the $F/B$ ratios, we also compared the radiated power of an emitter in the presence and absence of the antennas (both NRMA and NRDA), as shown in Fig. S13. First, the quantum emitter shows enhanced radiated power compared to free space, which is an advantage for experimental detection. Second, the emitter showed a higher radiated power with NRMA compared to NRDA, which is different to other configurations such as nanodisk[3] and nanostrip[4]. Third, the presence of the dielectric glass substrate does not affect the radiated power, but immersing the system in water can enhance it even further. Finally, increasing the length of AuNRs in NRDA seems to affect the radiated power mostly, compared to flattening tip or changing gap2.

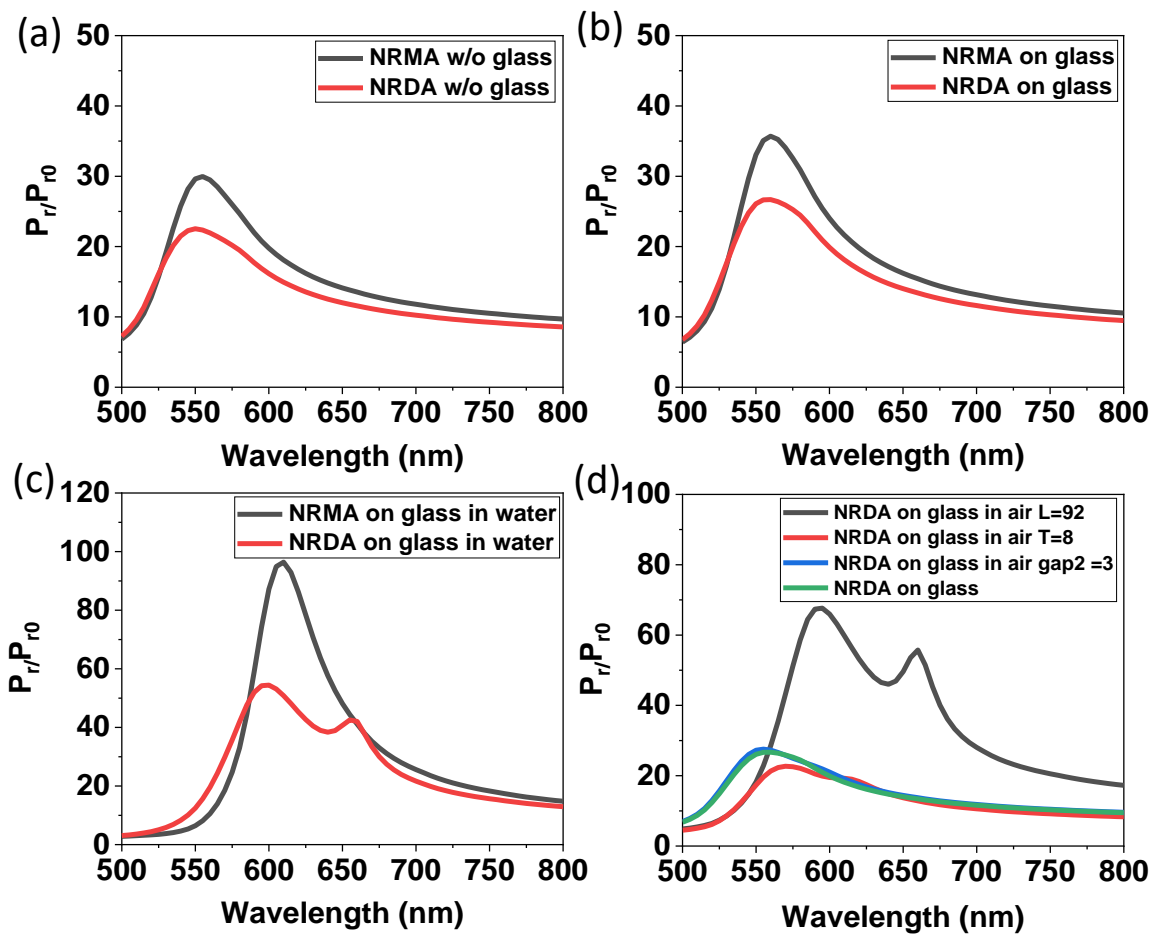

*Figure S13 Enhanced radiated power of antenna (NRMA or NRDA) compared to free dipole. NRDA and NRMA without glass substrate (a) and with glass substrate (b). (c) NRDA and NRMA with glass substrate in water. (d) NRDA with different size, shaded NR and gap2.*

# References

1. Kelly, K. L.; Coronado, E.; Zhao, L. L.; Schatz, G. C., The Optical Properties of Metal Nanoparticles: The Influence of Size, Shape, and Dielectric Environment. *The Journal of Physical Chemistry B* **2003,** *107* (3), 668-677.
2. Zhu, X.; Shi, H.; Zhang, S.; Dai, P.; Chen, Z.; Xue, S.; Quan, J.; Zhang, J.; Duan, H., Substrate-modulated directional far field scattering behavior of individual plasmonic nanorods. *Journal of Optics* **2020,** *22* (5).
3. Pakizeh, T.; Käll, M., Unidirectional Ultracompact Optical Nanoantennas. *Nano Letters* **2009,** *9* (6), 2343-2349.
4. Shen, H.; Lu, G.; He, Y.; Cheng, Y.; Gong, Q., Unidirectional enhanced spontaneous emission with metallo-dielectric optical antenna. *Optics Communications* **2017,** *395*, 133-138.